\documentclass{JHEP3}

\usepackage{amsmath}
\usepackage{amsfonts}
\usepackage{cite}
\usepackage{subfigure,epsfig}
\usepackage{psfrag}

\preprint{ZU-TH 16/10}

\title{Photon Radiation with MadDipole}

\author{T.\ Gehrmann$^a$,N.\ Greiner$^{b}$\\
$^a$ Institut f\"ur Theoretische Physik, Universit\"at Z\"urich,\\
Winterthurerstrasse 190, 8057 Z\"urich, Switzerland\\
$^b$
Department of Physics, University of Illinois at Urbana-Champaign,\\
1110 West Green Street, Urbana IL, 61801, USA}
\abstract{
We present the automation of a subtraction method
for photon radiation using the dipole formalism within the MadGraph framework. 
The subtraction terms
are implemented both in dimensional regularization and mass regularization for
massless and massive cases and non-collinear-safe observables are accounted for.
}

\begin{document}
\def\sub{{\mathrm{sub}}}
\def\gsub{g^{(\sub)}}
\def\Gsub{G^{(\sub)}}
\def\cGsub{{\cal G}^{(\sub)}}
\def\bcGsub{\bar {\cal G}^{(\sub)}}
\def\de{\delta}
\newcommand{\bra}[1]{\langle#1|}
\newcommand{\ket}[1]{|#1\rangle}
\newcommand{\nn}{\nonumber}
\newcommand{\zi}{\tilde z_i}
\newcommand{\zj}{\tilde z_j}
\newcommand{\sijk}{s_{ij,k}}
\newcommand{\yijk}{y_{ij,k}}
\newcommand{\vijk}{v_{ij,k}}
\newcommand{\viji}{v_{ij,i}}                                                    
\newcommand{\tvijk}{\tilde v_{ij,k}}
\def\aand{\!\!\!\!\!\!\!\!&&}
\newcommand{\eps}{\epsilon}
\newcommand{\alps}{\alpha_{\mathrm{s}}}
\newcommand{\CF}{C_{\mathrm{F}}}
\newcommand{\CA}{C_{\mathrm{A}}}
\newcommand{\TR}{T_{\mathrm{R}}}                                                
\newcommand{\Nc}{N_{\mathrm{c}}}       
\newcommand{\bV}       {{\bf V}}
\newcommand{\cV}       {{\cal V}}
\newcommand{\cK}       {{\cal K}}
\newcommand{\cT}       {{\cal T}}
\newcommand{\rd}{{\mathrm{d}}}
\newcommand{\tpij}{\widetilde p_{ij}}
\newcommand{\tpk}{\widetilde p_k}
\newcommand{\coll}{{\mathrm{coll}}}
\newcommand{\eik}{{\mathrm{eik}}}
\newcommand{\xija}{x_{ij,a}}
\newcommand{\M}{{\cal{M}}}
\newcommand{\DIS}{\ensuremath{{\mathrm{DIS}}}}
\newcommand\Oe[1]      {\ensuremath{\mathrm O(\ep^{#1})}}
\def\ep{\epsilon}
\newcommand\as{\alpha_{\mathrm{S}}}
\newcommand{\cut}{\mathrm{cut}}
\def\bom#1{{\mbox{\boldmath $#1$}}}
\def\KFS#1{K^{{#1}}_{\scriptscriptstyle\rm F\!.S\!.}}
\def\HFS#1{H_{{#1}}^{\scriptscriptstyle\rm F\!.S\!.}}
\def\Kab{\KFS{ab}}
\section{Introduction}
A precise theoretical description of cross sections at particle colliders is 
required for the reliable extraction of fundamental parameters, for 
indirect searches for new physics effects, and for a well-founded assessment 
of background reactions 
to new physics signatures and their uncertainties. This precision is achieved 
by inclusion of higher order perturbative corrections from QCD and 
from the electroweak theory. At hadron colliders, especially the 
inclusion of next-to-leading order (NLO) QCD appears to be mandatory 
for many studies. Much effort has been made in the recent past to 
automate NLO QCD calculations (see~\cite{Binoth:2010ra,Binoth:2010xt} 
for reviews). 

A priori, NLO electroweak corrections can be expected to be 
smaller than QCD corrections, owing to the smaller coupling constant.
In a variety of situations, this naive argumentation does however not apply. 
In QCD,  quarks and gluons can not be distinguished, and 
collinear gluon radiation off an outgoing parton can not be identified 
experimentally. Collinear photon radiation off a lepton leaves a 
signature in the detector and can thus be identified. As a result, 
the cancellation of photonic contributions in the virtual and real collinear 
contributions at NLO is incomplete, and the observable is 
non-collinear-safe~\cite{Dittmaier:2008md}. The final state photon 
selection cuts must be implemented consistently in the theoretical 
description, both at fixed order and in a parton shower 
framework~\cite{Hoeche:2009xc}. Depending on 
the treatment of photon radiation in the experimental analysis, 
the magnitude and shape of the electroweak corrections can 
differ substantially, due to the occurrence of large logarithmic terms in the
photon cuts, as discussed in detail for instance in the electroweak corrections 
to gauge boson production and decay~\cite{Dittmaier:2001ay,CarloniCalame:2006zq}. 
At centre-of-mass 
energies above the weak gauge boson masses, similar incomplete 
cancellations occur in the genuine weak corrections, giving rise 
to large logarithms in the gauge boson masses. 

NLO electroweak corrections for hadron collider observables have up to now been 
derived for gauge boson production~\cite{Dittmaier:2001ay,CarloniCalame:2006zq}, for 
gauge boson pair production~\cite{Accomando:2005ra}, for Higgs boson 
production and decay~\cite{Ciccolini:2007ec,Actis:2008ug,Bredenstein:2006ha,Ciccolini:2003jy,
Degrassi:2004mx,Degrassi:2005mc}, for top quark pair 
production~\cite{Kuhn:2006vh,Bernreuther:2008md}, single top production~\cite{Beccaria:2006dt}
 and for supersymmetric particle 
production~\cite{Hollik:2007wf,Hollik:2008vm,Germer:2010vn}. 
Much effort has been made recently towards 
electroweak corrections to final states including jets. For these 
final states, the experimental treatment of photons 
observed inside jets~\cite{Glover:1993xc} can yield 
non-collinear-safe configurations, 
resulting in potentially large corrections. A first calculation of 
this type was accomplished recently with the NLO corrections to $W^\pm$+jet 
final states~\cite{Denner:2009gj}. 

NLO electroweak corrections   
require two ingredients: the one-loop virtual corrections 
and the real radiation corrections with one extra final state particle. 
While the one-loop virtual corrections contain explicit infrared poles 
from the loop integration, the real radiation contributions develop
infrared singularities only when integrated over the real radiation phase 
space. To handle these real radiation singularities in a process-independent 
manner, subtraction formalisms have been developed initially in the 
context of QCD calculations: residue (or FKS) subtraction~\cite{Frixione:1995ms}
 and
variants thereof~\cite{Somogyi:2009ri}, dipole
subtraction~\cite{Catani:1996vz,Catani:2002hc} and antenna
subtraction~\cite{Kosower:1997zr,Campbell:1998nn,GehrmannDeRidder:2005cm,Daleo:2006xa}, all within dimensional regularization. These can in principle 
be carried over to electroweak calculations. However, some care has to be 
taken since most electroweak calculations use mass regularization.
The
dipole subtraction method has been extended to 
include photonic radiation in~\cite{Dittmaier:1999mb}.

Since NLO electroweak corrections are required for a substantial number of 
processes, their automation is very much desirable. 
Driven mainly by QCD calculations, the  automation of one-loop 
corrections for one-loop multi-parton amplitudes has seen enormous 
progress in the recent past, and  first
fully automated programs for the calculation of one-loop multi-parton
amplitudes are becoming available with the packages
CutTools~\cite{Ossola:2007ax,Ossola:2008xq},
BlackHat~\cite{Berger:2008sj}, Rocket~\cite{Giele:2008bc}, 
GOLEM~\cite{Binoth:2008uq} and Samurai~\cite{Mastrolia:2010nb}, as well as independent
libraries~\cite{Denner:2005nn}. 

Automation of the real radiation contributions 
is normally based on a leading order parton-level event generator 
framework such as
MadGraph/MadEvent
\cite{Stelzer:1994ta,Maltoni:2002qb,Alwall:2007st},
ALPGEN \cite{Mangano:2002ea},
 CompHEP/CalcHEP
\cite{Boos:2004kh}/\cite{Pukhov:2004ca}, SHERPA
\cite{Gleisberg:2003xi,Gleisberg:2008ta}, HELAC \cite{Cafarella:2007pc} or 
WHIZARD~\cite{Kilian:2007gr}.
The generation of dipole terms for
subtracting the singular behaviour from the real radiation
subprocesses has been automated in various event generators: in the
SHERPA framework \cite{Gleisberg:2007md}, the TeVJet framework
\cite{Seymour:2008mu}, the HELAC framework~\cite{Czakon:2009ss} and in
the form of independent libraries~\cite{Hasegawa:2009tx,Hasegawa:2010dz} interfaced to
MadGraph. The MadDipole package~\cite{Frederix:2008hu} provides an
implementation within MadGraph. An implementation of the residue
subtraction method is also available within
MadGraph~\cite{Frederix:2009yq}. For a full NLO calculation, the dipole terms have to be integrated
over the dipole phase space, and added with the virtual corrections to
obtain the cancellation of infrared singularities. The 
HELAC implementation~\cite{Czakon:2009ss} and the MadDipole 
package~\cite{Frederix:2010cj} 
provide these integrated
dipole terms including all masses and possible phase-space
restrictions. 

It is the purpose of this work to extend the MadDipole package towards 
electroweak NLO corrections by automatically generating the 
unintegrated and integrated dipole subtraction terms for 
photonic radiation, as defined in~\cite{Dittmaier:1999mb}, and 
including non-collinear-safe configurations~\cite{Dittmaier:2008md}. \\This 
paper is structured as follows: in Section~\ref{sec:reg}, we review the 
differences between mass regularization and dimensional regularization, and 
their implications for the dipole terms. The treatment of photonic effects 
in the parton distributions is described in Section~\ref{sec:pdf}, and 
the dipole terms for non-collinear-safe observables are introduced in 
Section~\ref{ncs}. The implementation of photon radiation in MadDipole 
is documented in Section~\ref{sec:imp}, and checks of it are 
described in Section~\ref{sec:check}. 

\section{Dimensional regularization vs.\ mass regularization}
\label{sec:reg}
Infrared singularities are an inherent property of quantum field theories. In the dipole formalism
this becomes exlicit when integrating over the one particle phase space to obtain the integrated
subtraction terms. As the integration over this phase space is divergent one needs a prescription
to parametrize the singularities. For QCD calculations dimensional regularization  is the most
widely
used method. The singularities are present in poles of the form $1/\epsilon^2$ and $1/\epsilon$.
Electroweak calculations are mostly done using mass regularization where the 
actual masses of the fermions and a hypothetical mass parameter of the photon are 
used as infrared cut-offs.   
In this case the singularities show up as logarithms of the mass parameter.\\ 
In our implementation we offer both possibilities and the user can select
 which method is used.
The implementation within dimensional regularization 
works in the same way as for the QCD version of
MadDipole, described in~\cite{Frederix:2008hu,Frederix:2010cj}.\\
In handling photon radiation, the main differences are in the initial state radiation 
collinear singularities, which lack counterparts in the virtual corrections. 
In QCD these singularities are absorbed by a redefinition of the parton distribution 
functions.  Integrated subtraction terms for
processes with initial state partons contain distributions in the momentum fraction $x$ of 
the parton entering the hard scattering process (for details see
\cite{Catani:1996vz,Catani:2002hc}). A priori the reduced matrix element also depends on $x$,
a feature that bears disadvantages for the numerical implementation: 
 the matrix element has to be evaluated
at $x\ne1$ and at $x=1$. In the presence of a parton distribution function this problem can be
circumvented by a transformation of variables, thus shifting the dependence of this variable into
the parton distribution function. The details and how this is done in MadDipole is described 
in \cite{Frederix:2010cj}.\\ \\
The major new aspects in the implementation of photon radiation are the following:\\
{\bf 1.} Initial state radiation singularities 
in electroweak processes are regulated by the masses of leptons 
entering the hard process, resulting in potentially large mass logarithms. In contrast to 
the divergencies absorbed into the parton distribution functions, these mass logarithms have 
physical significance. Consequently, for 
electroweak corrections to processes with leptons in the initial state, 
the user has to choose mass regularization as the appropriate
scheme. Furthermore, it is not possible to remove the $x$-dependence from the 
hard scattering matrix element by a variable redefinition, which thus has to be evaluated 
up to three times: at $x=1$, and at the values of $x$ corresponding to the initial state 
radiation off each of the two incoming momenta.\\
{\bf 2.} For configurations containing only a collinear but not a soft divergence, 
the summation over the dipole spectator momenta is obsolete, since these configurations 
can be accounted for by a single dipole term. In the QCD
MadDipole implementation, this summation is nevertheless carried out in order to 
facilitate the colour management. In the electroweak MadDipole implementation, this summation 
is omitted. For the convenience of the user, all possible spectator terms are generated, and 
can be invoked manually. 

\section{QED corrections to parton distribution functions}
\label{sec:pdf}
For processes with incoming partons, initial state collinear singularities 
 are absorbed into the parton distribution functions (PDFs). 
 In dimensional regularization this collinear
singularity is of the form
\begin{equation}
 \frac{(4\pi\mu^2)^{\epsilon}}{\Gamma(1-\epsilon)}\frac{1}{\epsilon}P_{ji}(x)\;,
\label{collsing}
\end{equation}
 where $P_{ji}$ denotes the Altarelli-Parisi splitting function for the 
 $j\to i$ splitting.\\
 For example, if one considers gluon radiation off an incoming quark, the 
 absorption of the collinear singularity of \ref{collsing} into the quark PDF  leads to
 \begin{equation}
 q(x,\mu_F) = q_0(x)+ \frac{\alpha_s}{2\pi}\int \limits_x^1\frac{dy}{y}q_0(y)\left[\frac{-1}{\Gamma(1-\epsilon)}
\left(\frac{4\pi\mu^2}{\mu_F^2}\right)^{\epsilon}\frac{1}{\epsilon}P_{qq}\left(\frac{x}{y}\right)
 + C_{qq}\left(\frac{x}{y}\right)\right]\;,
\end{equation}
where $\mu_F$ is the factorization scale. 
The value of the coefficient $C_{qq}\left(\frac{x}{y}\right)$ defines the subtraction scheme.
The $\overline{MS}$ scheme which is mainly used in QCD just absorbs the singular terms and
some trivial constants, so 
\begin{equation}
C_{qq}^{\overline{MS}}(x) = 0\;.
\end{equation}
For the subtraction of photon radiation, 
the equivalent expressions for QED corrections using mass regularization are needed.
This can be derived by calculating the integrated subtraction terms both in dimensional and
mass regularization and comparing the results.
A detailed derivation can be found in 
\cite{Dittmaier:2009cr,Diener:2005me,Ciccolini:2007ec}, so we only quote the results 
here.
According to Eqns.~(3.19) and (3.20) of ~\cite{Dittmaier:2009cr}, to include
inital state radiation at NLO QED of a parton $a$ out of a hadron $h$ with longitudinal momentum fraction $x$,
the parton distribution functions are modified by
\begin{eqnarray}
  f^{(h)}_{q/\bar q}(x) \rightarrow f^{(h)}_{q/\bar q}(x,\mu_{\mathrm{F}}^2) 
  &-&\; \frac{\alpha\,Q_q^2}{2\pi} \; \int^1_x \frac{\rd z}{z}\; 
  f^{(h)}_{q/\bar q}\left(\frac{x}{z},\mu_{\mathrm{F}}^2\right) \\
  && \times \left\{\ln\biggl(\frac{\mu_{\mathrm{F}}^2}{m_q^2}\biggr) \; 
  \left[P_{ff}(z)\right]_+ - \left[P_{ff}(z)\;(2\ln(1-z)+1)\right]_+ 
  + C^\DIS_{ff}(z)\; \right\} \nn \\
  &-&3 \frac{\alpha\,Q_q^2}{2\pi} \;\int^1_x \frac{\rd z}{z} \; 
  f^{(h)}_\gamma\left(\frac{x}{z},\mu_{\mathrm{F}}^2\right) \; 
  \left\{\ln\biggl(\frac{\mu_{\mathrm{F}}^2}{m_q^2}\biggr) \; P_{f\gamma}(z) +
  C^\DIS_{f\gamma}(z) \;\right\}, \nn \\
  f^{(h)}_{\gamma}(x) \rightarrow f^{(h)}_\gamma(x,\mu_{\mathrm{F}}^2) &-&\; \frac{\alpha\,Q_q^2}{2\pi}
  \sum_{a=q,\bar{q}} \int^1_x \frac{\rd z}{z} \;
  f^{(h)}_a\left(\frac{x}{z},\mu_{\mathrm{F}}^2\right) 
\\
  && \qquad \times \left\{\ln\biggl(\frac{\mu_{\mathrm{F}}^2}{m_q^2}\biggr) \;
  P_{\gamma f}(z) 
  - P_{\gamma f}(z)\,(2\ln z+1)
  + C^\DIS_{\gamma f}(z) \;\right\}\nonumber\;,
\end{eqnarray}
with the splitting functions
\begin{equation}
  P_{ff}(z)   = \frac{1+z^2}{1-z}\,, \qquad
  P_{f \gamma}(z)= z^2 + (1-z)^2\,, \qquad
  P_{\gamma f}(z)= \frac{1 + (1-z)^2}{z}\;.
\end{equation}
The only set of PDFs which include both QCD and QED NLO corrections is
the MRST2004QED \cite{Martin:2004dh}. As has been pointed out in \cite{Diener:2005me},
a consistent use of this set of PDFs requires the use of the DIS scheme for QED
calculations. The coefficient functions for the DIS scheme are given by
\begin{eqnarray}
    C^{\DIS}_{ff}(z) &=& \left[ \; P_{ff}(z) \; 
    \biggl( \; \ln
    \biggl( \frac{1-z}{z} \biggr) - \frac{3}{4} \; \biggr)
    + \frac{9+5z}{4} \; \right]_+ \;,\nn\\
    C^{\DIS}_{f\gamma}(z) &=&
    P_{f\gamma}(z)\;\ln\biggl(\frac{1-z}{z}\biggr)
    - 8z^2 + 8z -1 \;,\nn\\
    C^{\DIS}_{\gamma f}(z) &=& - C^{\mathrm{DIS}}_{ff}(z) \;. 
  \label{CDIS}
\end{eqnarray}
It should be noted  
that the integrated subtraction terms as they appear
in MadDipole are different for processes with incoming PDF  and for processes without.  
The introduction of the PDF does not only affect the logarithmically
divergent terms,  where the mass of the particle is replaced by the factorization
scale, but  also finite parts are different due to the coefficient functions present in the PDFs.
Whether a PDF is present or not is determined automatically by MadDipole in the case
of leptons or QCD particles. Any QCD particle 
(quark, anti-quark, gluon) is assumed to be a constituent of a hadron whereas this is certainly
not true for leptons. A special case is the 
incoming photon, because it may come out of a hadron  but
could as well be a free particle. Here the user has to choose whether to use a
PDF or not (see section \ref{sec:imp}).

\section{Non-collinear-safe observables}
\label{ncs}
Final state collinear photons can often be distinguished from other final state particles. For 
example, the energy fractions in a collinear muon-photon system can be disentangled 
by comparing calorimetric and tracking information, and highly energetic photons inside QCD 
jets can be identified and vetoed. Therefore, in 
electroweak calculations a collinear photon is not necessarily treated 
fully inclusively \cite{Dittmaier:2008md}.
Consequently, the integration range within the cone of a photon radiated off a charged particle
is not the full range of the momentum fraction of the charged particle. This momentum
fraction is parametrized by the variable $z_{ij}$ for a final-final constellation and $z_{ia}$ 
for a final-inital one. Treating the photon not fully inclusively is now equivalent to imposing
a cut on this variable. In this case, the collinear singularity associated 
with the photon emission off a massless fermion is not cancelled, rendering 
the observable non-collinear-safe. In analogy to the factorization of 
initial state collinear singularities into parton distributions, 
this singularity is absorbed into 
a photon fragmentation function~\cite{Koller:1978kq,Glover:1993xc}. 
An alternative photon isolation procedure has been proposed in the form of a dynamical cone-size~\cite{Frixione:1998jh},
which would in principle shield the collinear singularity and eliminate the need for the photon
fragmentation function. In practise, this criterion turns out to be difficult
to implement for a finite-resolution detector. \\
The treatment of fragmentation processes in the dipole formalism was 
first developed in~\cite{Catani:1996vz} for dimensional regularization, and 
extended to mass regularization in~\cite{Dittmaier:2008md}. We largely
follow the latter paper for notation and conventions and
refer to this reference for
further details. Here we collect those results which were important for our implementation.\\ \\
For observables with identified final state photons, 
the real emission matrix element and the unintegrated subtraction 
terms are multiplied by 
a step function that cuts off the integration at a given value for $z$. One therefore has
\begin{eqnarray}
 && \int\rd\Phi_1\, \Biggl[
\sum_{\lambda_\gamma}|\M_1|^2
\Theta_{\cut}(p_f,k,p_{f'},\{k_n\})
\nn\\
&& \hspace{3.5em} {}
-\sum_{f\ne f'} |\M_{\sub,ff'}|^2
\Theta_{\cut}\Bigl(z_{ff'} \tilde p_f^{(ff')},(1-z_{ff'})\tilde p_f^{(ff')},
\tilde p_{f'}^{(ff')},\{k_n\}\Bigr)
\Biggr].
\label{eq:thetacutme}
\end{eqnarray}
The integrated dipole terms are thus not integrated over $z$, but remain functions of this 
variable. \\ \\
The MadDipole implementation usually allows for the introduction of a phase space 
cut, $\alpha$, 
to ensure that dipoles are subtracted only in the vicinity of the singular regions. 
In our implementation, non-collinear-safe observables can only be calculated 
with all $\alpha$-parameters set to one, since including the $\Theta$-function related to the 
$\alpha$-parameter in them would lead to very involved phase space constraints, which 
prevent the analytical integration of the subtraction terms. 

For non-collinear safe-observables, the emitter is always in the final state, while the 
spectator can be in final or initial state. Both cases were derived in~\cite{Dittmaier:2008md} 
and are summarized below. 

\subsection{final-final}
In the final-final case the integrated splitting function in mass regularization is given by
\begin{equation}
 \bcGsub_{ij,\tau}(P_{ij}^2,z_{ij}) = 
\frac{\bar P_{ij}^4}{2\sqrt{\lambda_{ij}}}
\int_{y_1(z_{ij})}^{y_2(z_{ij})} \rd y_{ij}\, (1-y_{ij})\,
\gsub_{ij,\tau}(p_i,p_j,k)\;.
\end{equation}
For an appropriate numerical treatment the soft singularity at $z=1$ is regularized by 
introducing a plus distribution for $z$,
\begin{equation}
 \bcGsub_{ij,\tau}(P_{ij}^2,z) =
\Gsub_{ij,\tau}(P_{ij}^2) \delta(1-z) 
+ \left[\bcGsub_{ij,\tau}(P_{ij}^2,z)\right]_+\;.
\end{equation}
The integration boundaries can be derived from the phase space, for details see Appendix B of\cite{Dittmaier:1999mb}.
In the limit $m_i\to 0$ and $m_j, m_{\gamma} =0$ one finds
\begin{equation}
y_1(z) = \frac{m_i^2(1-z)}{P_{ij}^2 z}, \quad 
y_2(z) = 1.
\end{equation}
Performing the integration leads to
\begin{eqnarray}
\bcGsub_{ij,+}(P_{ij}^2,z) &=&
P_{ff}(z)\,\biggl[ \ln\biggl(\frac{P_{ij}^2 z}{m_i^2}\biggr)-1\biggl]
+(1+z)\ln(1-z), \nn\\ 
\bcGsub_{ij,-}(P_{ij}^2,z) &=& 1-z,
\end{eqnarray}
with
\begin{equation}
P_{ff}(z) = \frac{1+z^2}{1-z}\;.
\end{equation}
The case where $m_j$ is not zero is more involved however there is no conceptual difference to the case
of a massless spectator so we do not quote the result here. It can be found in Appendix A of \cite{Dittmaier:2008md}.

The derivation of the corresponding expressions in dimensional regularization 
starts from the dipole splitting function
\begin{equation}
 \frac{<{\bom V}_{ij,k}({ z}_i;y)>}{8 {\pi} \as
\mu^{2\ep}} = 
\frac{1}{(1-y_{ij,k})} \biggl[
\frac{2}{1-z_{i}(1-y_{ij,k})}-1-z_{i}-\eps(1-z_{i}) \biggr]\; ,
\end{equation}
which integrates to 
\begin{equation}
 {\cal V}(\epsilon,z) = -\frac{1}{\epsilon}P_{ff}(z)+P_{ff}(z)\ln(z) +\frac{z^2+3}{1-z}
 \ln(1-z)+(1-z)\;,
\end{equation}
In the case where the spectator is massive, the splitting function takes the form
\begin{equation}
 \frac{<{\bom V}_{ij,k}({z}_j;y)>}{8 {\pi} \as
\mu^{2\ep}} = 
\frac{1}{R_{ij}(y)} \biggl[
\frac{2}{1-z_{j}(1-y_{ij,k})}-\frac{\tilde{v}_{ij,k}}{v_{ij,k}}(1+z_{j}+\eps(1-z_{j})) \biggr]\; ,
\end{equation}
integrating to 
\begin{flalign}
  {\cal V}(\epsilon,z) &=-\frac{P_{ff}(z)}{\epsilon}+\frac{1}{z-1}\left\{z^2 
\left(-\log \left(8( c_1- c^2)\right)+\log (4 (c (c+c_2)- c_1 y_2))+\log (c_3)\right)
\right.\nn \\&\left.
+\log   \left(c_1-c^2\right)-\log (c (c+c_2)-c_1 y_2)-2 \left(z^2-1\right) \log (c)-\log (c_3)
\right.\nn \\&\left.
+2 \log ((y_2-1)
   z+1)-2 \log (1-z)+\log (2)\right\}+P_{ff}(z)\log (-c y_2 (z-1) z)-z+1,
\end{flalign}
where we used the abbreviations
\begin{equation}
 c=1-\mu_k^2\;,\quad c_1=1-\mu_k^4\;,\quad c_2=\sqrt{c^2(y_2^2+1)-2c_1y_2}\;,\quad
c_3=c_1-c(c\ y_2+c_2)
\end{equation}
and
\begin{equation}
 y_2(z)=\left[ \xi(z)+1+\sqrt{\xi(z)(\xi(z)+2)}\right]^{-1}\;, \qquad \xi(z)=\frac{\mu_k^2}{2z(1-z)}
\end{equation}
to compactify the result.

\subsection{final-initial}
For an initial state spectator the subtraction terms depend on the variable $x_{ia}$ and
are divergent for $x_{ia} \to 1$. This is taken into account by splitting off the divergent
parts using the $(+)$-distribution. The correction terms for non-collinear-safe observables now
come along with an additional distribution in $z$ so the the full result reads as a combination
of the two.
\begin{eqnarray}
\lefteqn{
-\frac{\bar P_{ia}^2}{2} \int_{0}^{x_1} \rd x\, 
\int_{z_1(x)}^{z_2(x)} \rd z \, \gsub_{ia,\tau}(p_i,p_a,k)
\cdots } &\qquad&
\nn\\*
&=& \int_{0}^{1} \rd x\, \int_{0}^{1} \rd z\, \left\{
\Gsub_{ia,\tau}(P_{ia}^2) \, \de(1-x) \, \de(1-z) 
+ \left[\cGsub_{ia,\tau}(P_{ia}^2,x)\right]_+ \de(1-z) \right.
\nn\\
&& \hspace{6em}\left. {}
+ \left[\bcGsub_{ia,\tau}(P_{ia}^2,z)\right]_+ \de(1-x)
+ \Bigl[\bar{g}^{\mathrm{(sub)}}_{ia,\tau}(x,z)\Bigr]^{(x,z)}_+ 
\right\} \cdots,
\label{eq:Gia2}
\end{eqnarray}
with
\begin{eqnarray}
 \bar{g}^{\mathrm{(sub)}}_{ia,+}(x,z) &=&
\frac{1}{1-x}\left(\frac{2}{2-x-z}-1-z\right),
\qquad
\bar {g}^{\mathrm{(sub)}}_{ia,-}(x,z) = 0,
\nn\\
\bcGsub_{ia,+}(P_{ia}^2,z) &=&
P_{ff}(z)\biggl[\ln\biggl(\frac{-P^2_{ia}z}{m_i^2}\biggr)-1\biggr]
-\frac{2\ln(2-z)}{1-z}+(1+z)\ln(1-z),
\nn\\
\bcGsub_{ia,-}(P_{ia}^2,z) &=& 1-z.
\end{eqnarray}

The first contribution, $\bar{g}^{\mathrm{(sub)}}_{ia,\tau}(x,z)$ is identical
in mass regularization and dimensional  regularization.\\
To evaluate the  equivalent expression of $\bcGsub_{ia,\tau}(P_{ia}^2,z)$ in dimensional
regularization we use the splitting function and phase space given in Eqns.~(5.39) and
(5.48) of~\cite{Catani:1996vz} and find
\begin{equation}
 {\cal V}(\epsilon,z)=-\frac{P_{ff}(z)}{\epsilon}+P_{ff}(z)\log(z)+\frac{z^2+3}{1-z}\log(1-z)
-2\frac{\log(2-z)}{1-z}+1-z\;.
\end{equation}

\subsection{Factorization of the collinear final state singularity}
The integrated dipole factors contain a collinear singularity. For fully 
inclusive observables, the dipole factors would be integrated over the 
photon momentum fraction $z$ and added with the virtual photon corrections,
rendering the result finite. Since non-collinear-safe observables
explicitly require (or veto) the final state photon, these cancellations no 
longer take place. In the case of photon radiation off massive 
fermions, this results in a large logarithm in the fermion mass, which 
is present in the above mass-regularization results. For massless fermions, 
especially for light quarks, this turns into a collinear singularity, and 
one has to include a contribution from the 
fragmentation of a final state fermion into a photon-fermion system 
of low invariant mass. The quark-to-photon 
fragmentation function (which depends on the photon momentum 
fraction $(1-z)$) is 
an empirically determined quantity~\cite{Koller:1978kq,Glover:1993xc,Buskulic:1995au,GehrmannDeRidder:1997wx}.  
It contains a counterterm for the collinear singularity, whose 
explicit form
depends on the regularization scheme. The counterterm is accounted 
for by adding the bare fragmentation function to the integrated 
dipole terms.
In mass regularization, the bare fragmentation function
reads~\cite{Denner:2010ia}:
\begin{eqnarray}
\label{eq:DfragMR}
D_{q\rightarrow\gamma}^{\mathrm{bare,MR}}(1-z)&=&
D_{q\rightarrow\gamma}(1-z,\mu_F)+
\frac{\alpha Q_q^2}{2\pi}                    
P_{ff}(z)\left[\ln\left(
\frac{m_q^2}{\mu_F^2}(1-z)^2
\right)+1
\right],
\end{eqnarray}
while the corresponding expression in dimensional regularization 
is~\cite{Glover:1993xc}: 
\begin{eqnarray}
\label{eq:DfragDR}
D_{q\rightarrow\gamma}^{\mathrm{bare,DR}}(1-z)&=&D_{q\rightarrow\gamma}(1-z,\mu_F)+
\frac{\alpha Q_q^2}{2\pi}                    
\frac{1}{\epsilon}\left(
\frac{4\pi\mu^2}{\mu_F^2}\right)^\epsilon\frac{1}{\Gamma(1-\epsilon)}
P_{ff}(z)\,.
\end{eqnarray}
It can be seen that the sum of the integrated dipole terms and the 
bare fragmentation function is independent on the regularization scheme. 
The scale-dependent non-perturbative fragmentation function 
$D_{q\rightarrow\gamma}(1-z,\mu_F)$
has to be determined from data.\\
Our implementation offers the  possibility to include the fragmentation
function. For details see section \ref{sec:imp}.

\section{Implementation and how to use it}
\label{sec:imp}
The installation and running of the new package is identical to the 
QCD version:

\begin{itemize}
\item[1.] Download the MadDipole package (version
  4.4.35 or later), \texttt{MG\_ME\_DIP\_V4.4.??.tar.gz}, from one of
  the MadGraph websites, {\it e.g.},
  \texttt{http://madgraph.hep.uiuc.edu/}.
\item[2.] Extract and run \texttt{make} in the \texttt{MadGraphII}
  directory.
\item[3.] Copy the \texttt{Template} directory into a new directory,
  {\it e.g.}, \texttt{MyProcDir} to ensure that you always have a
  clean copy of the Template directory.
\item[4.] Go to the new \texttt{MyProcDir} directory and specify your
  process in the file \texttt{./Cards/proc\_card.dat}.  This is the
  $(n+1)$-particle process you require the subtraction term for. In this
 file there is also a variabe \texttt{DipolePhotonPdf} which is set to
  \texttt{TRUE} by default. If so it is assumed that initial state photons
 are coming from a hadron and therefore a PDF is used. 
\item[5.] Running \texttt{./bin/newprocess\_qed} generates the code for
  the $(n+1)$-particle matrix element and for all dipole terms and
  their integrated versions. After running this you will find a newly
  generated directory \texttt{./SubProcesses/P0\_yourprocess} ({\it
    e.g.}, \texttt{./SubProcesses/P0\_e+e-\_uuxa}) which contains all
  required files.
\end{itemize}

In the \texttt{./SubProcesses/P0\_yourprocess} directory all the files
relevant to that particular subprocesses are generated. In particular
this includes the $(n+1)$ particle matrix elements in the file
\texttt{matrix.f} and the dipoles in the files \texttt{dipolqed???.f},
where \texttt{???} stands for a number starting from
\texttt{001}. Furthermore the directory has two files,
\texttt{dipolsumqed.f} and \texttt{intdipolesqed.f}, where the sum of the
dipoles and their integrated versions are calculated, respectively.\\ \\
The routines contained in \texttt{intdipolesqed.f} which calculate the integrated
subtraction terms have a slightly different syntax compared to the QCD version so we
give some more details here. As for QCD the file contains two routines
\begin{quote}
 \texttt{intdipolesqed(P,\ P1 ,\ P2 ,\ X,\ Z,\ Zi,\ PSWGT\, EPSSQ,\ EPS,\ FIN)} \textrm{ and}\\
 \texttt{intdipolesqedfinite(P,\ P1 ,\ P2 ,\ X,\ Z,\  PSWGT\, EPSSQ,\ EPS,\ FIN)}.
\end{quote}
The difference is that the routines need three phase space points as input variables.
This is due to the problem explained in section 1, that in the absence of a PDF the
reduced matrix element depends on the longitudinal momentum fraction $x$ 
(here named as \texttt{Z}).
Therefore the phase space point \texttt{P(0:3,nexternal)} denotes the point in
an $n$-particle phase space with $x=1$. The next variable, \texttt{P1(0:3,nexternal)}
denotes the phase space point with an initial state radiation from the first incoming
particle, which means its momentum is given by $x\cdot p_1$ where $p_1$ is the full
first momentum as it appears in the first phase space point. The third phase space point
is the same as second but with the radiation off the second incoming particle.\\
These phase space points have to be provided by the user. In the case where there are PDFs
present for both incoming particles we use the same method as in QCD, namely shifting the
dependence of the matrix element on the momentum fraction $x$ into the PDF. In that case
only the first phase space point \texttt{P(0:3,nexternal)} has to be provided, the others
are not used.\\ \\
The second difference to QCD is the appearence of the variable \texttt{Zi}. It is only needed
for non-collinear-safe observables (see section \ref{ncs}) and denotes the integration
variable $z$ as defined in that context. We assume that the integration over \texttt{Zi}
is carried out numerically and therefore this variable has to be provided by the user.\\ \\
Several important parameters are included in the file \texttt{dipole.inc} which need to
be set by the user.\\
The first parameter decides which regularization scheme should be used. It can be
either \texttt{DREG} or \texttt{MREG} which stands for dimensional regularization and
mass regularization respectively. Note that in the case where the reduced matrix element
contains leptons in the initial state, one has to choose mass regularization.\\
The second parameter is a logical variable \texttt{photonpdf} which determines whether
a PDF should be used for initial state photons. Whether this parameter is true or
false per default is determined by the choice the user made in \texttt{param\_card.dat}.
If the according variable in the parameter card is set to true, this variable
in \texttt{dipole.inc} will be set to true automatically when the code is generated.
So for consistency this parameter should not be changed by the user.
The next variable is a real array \texttt{massint} which contains the particle masses and
which have to be given by the user. More precisely, what is required are the masses of the
particles as they appear in the integrated subtraction terms. As an example we can take
$e^{+}e^{-}\to t \bar{t}\gamma$. For the real emission matrix element and for the unintegrated
dipoles one would usually treat the electron as a massless particle whereas the top-quark
would be treated as a massive particle. The integrated subtraction terms however require
an electron mass to regulate the infrared singularities. These regulator masses are declared 
in the array. So in this example one would fill the first two entries with
the electron mass, and the next two entries would be the top mass. A photon should be treated
massless in this array, so the last entry would be zero. All  units are in $GeV$.
Per default all entries are set to one to ensure a finite result.\\
The next parameter is the photon mass
parameter, which is also needed to regulate logarithmic
divergencies. By default it is set to unity so that all logarithms containing the 
photon mass vanish. \\
The following parameter is a logical variable \texttt{ncs}. If set to true it calculates the
correction terms for non-collinear safe observables. This requires of course that
the user has to provide appropriate values for the variable \texttt{Zi} described
above.\\
The last parameter is given by a logical variable \texttt{photonfrag} which is
set to true by default. If set to true the second terms on the right hand side of
Eqns.\ (\ref{eq:DfragMR}/\ref{eq:DfragDR}) are added to the integrated dipoles and a
template function is called which can be used by the user to provide the non-perturbative
parts of the fragmentation function, given by the first term on the right hand side
of Eqns.\ (\ref{eq:DfragMR}/\ref{eq:DfragDR}). This template function is described in
section \ref{sec:ncsimp}.

\subsection{Implementation of non-collinear-safe case}
\label{sec:ncsimp}
The calculation of non-collinear-safe observable requires to introduce additional
cuts which act on the matrix element as well as on the subtraction terms, as in Eq.~(\ref{eq:thetacutme}).
In MadDipole these cuts can be imposed using the predefined cut routines, no additional 
routines are introduced. The correct handling
of the $\Theta$-function has to be done by the user.\\
Also the integrated subtraction terms receive additional cuts in form of a $\Theta$-function.
For the final-final case, the integrated subtraction terms are therefore of the form
\begin{eqnarray}
\label{eq:ffintz}
\int\rd\Phi_1\,|\M_{\sub,ij}(\Phi_1;\kappa_i)|^2
&=& -\frac{\alpha}{2\pi} Q_i\sigma_i Q_j\sigma_j  \,
\int\rd\tilde\Phi_{0,ij}\,
\int_0^1 \rd z\, 
\nn\\[.3em]
&& {}\times
\left\{ \Gsub_{ij,\tau}(P_{ij}^2) \de(1-z) 
+ \left[\bcGsub_{ij,\tau}(P_{ij}^2,z)\right]_+ \right\}
\\[.3em]
&& {}\times
|\M_0(\tilde p_i,\tilde p_j;\tau\kappa_i)|^2 \,
\Theta_{\cut}\Bigl(p_i=z\tilde p_i,k=(1-z)\tilde p_i,\tilde p_j,\{k_n\}\Bigr),
\nn
\end{eqnarray}
The parts that are $z$-dependent are the $(+)$-distribution and the $\Theta$-function. This means
that this $\Theta$-function regularizes the singularity contained in the $(+)$-distribution at
$z=1$. We therefore need to evaluate the step function at two points, $z \ne 1$ and $z=1$. 
As this can not be done within the predefined cut routines, we introduce an additional 
logical variable
\begin{itemize}
 \item \texttt{thetacut(P,zi)}
\end{itemize}
 which can be either true or false.
Its arguments are the phase space point and the variable $z$. As MadDipole uses a variable $z$
already to denote the momentum fraction in the context of initial state radiation we rename this
variable to \texttt{zi}. In the same way
as for the 'normal' cut routine MadDipole provides an empty dummy routine in which the user
can specify the desired cuts. MadDipole then takes care that it is evaluated at the two different
points and combined with the subtraction terms according to (\ref{eq:ffintz}). The template 
of this function is included in the file \texttt{intdipolesqed.f}.\\ \\
In the same file we also provide a template that can be used for calculating the
non-perturbative parts of the quark-to-photon fragmentation function as described in
the previous section. The template is given by the real function
\begin{itemize}
 \item \texttt{DFRAG(zi)}
\end{itemize}
where the only input variable is $z$. Here the user should provide a call to the set of
fragmentation functions of his choice. The result will then be added automatically to
the integrated dipole terms. Per default, this routine returns zero so no non-perturbative
terms are added.

\section{Checks}
\label{sec:check}
\begin{figure}[t]
\label{alphadep}
  \begin{center}
    \mbox{
      \subfigure[]{
        \epsfig{file=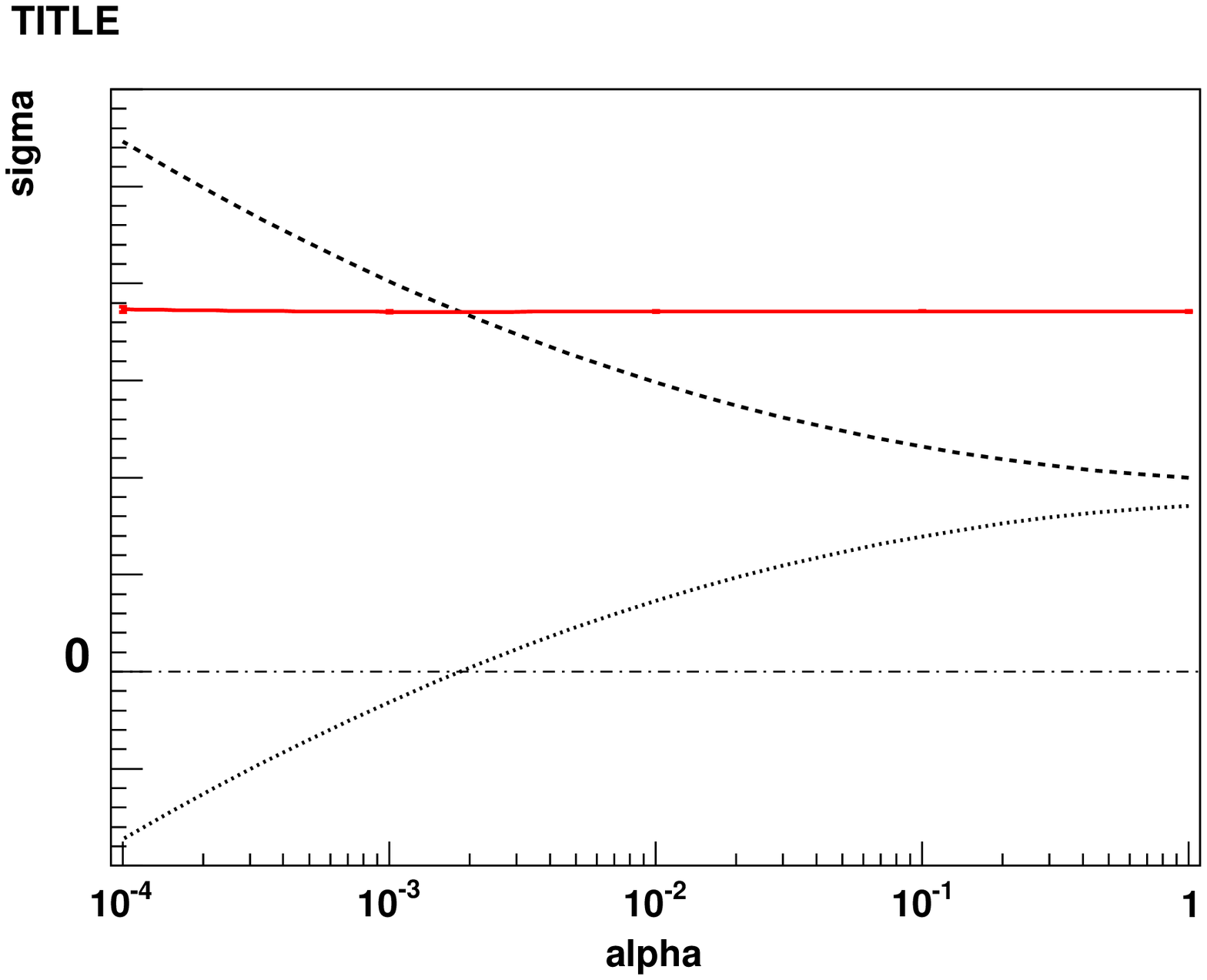, width=0.48\textwidth}
      }
      \subfigure[]{
        \epsfig{file=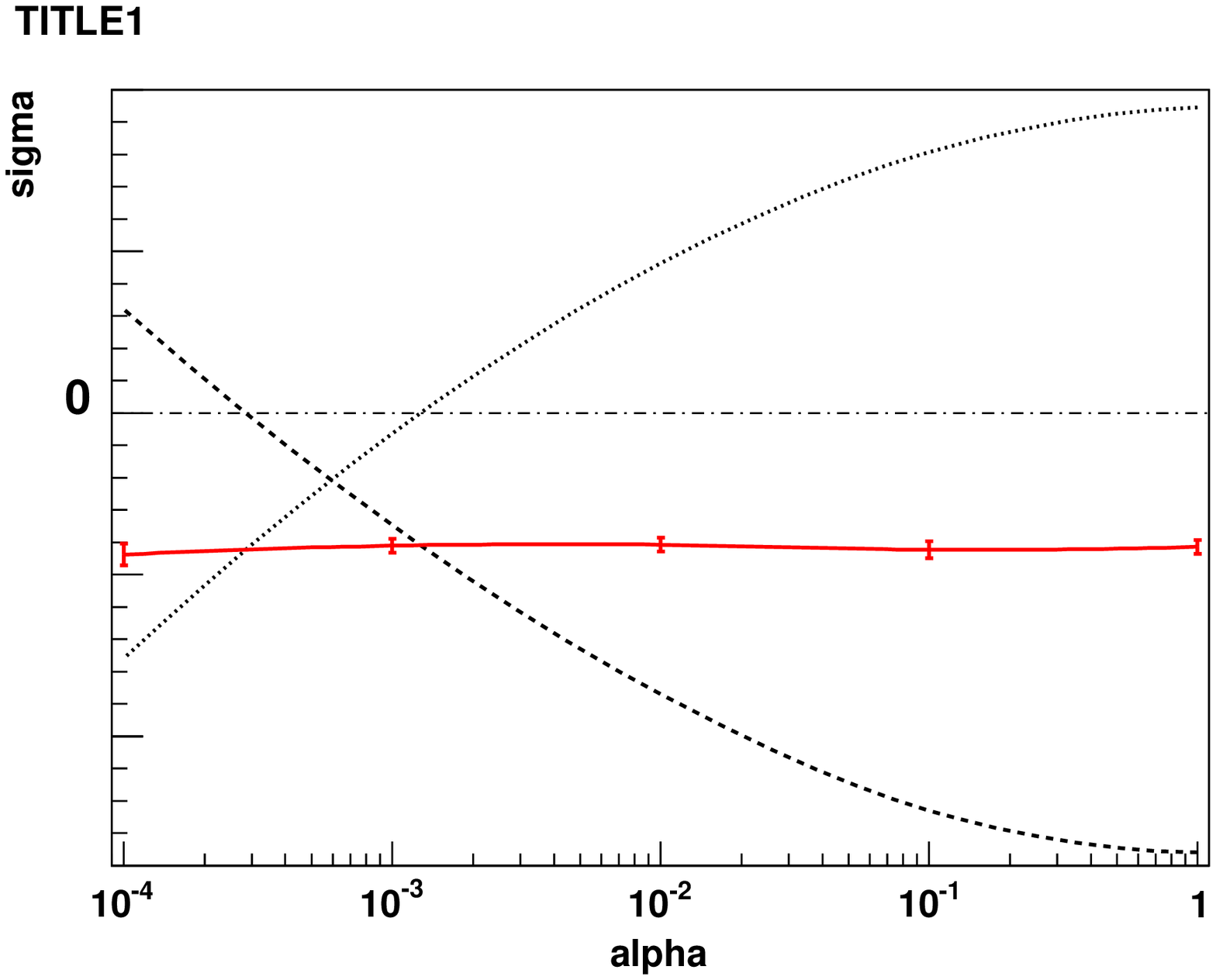, width=0.48\textwidth}
      }
    }
  \end{center}
  \vspace{-20pt}
  \caption{Dependence of the cross section on the cut parameter $\alpha$.
 All four $\alpha$-parameters (final-final, final-initial, initial-final, initial-initial)
are varied simultaneously. The  dashed line 
 denotes the real emission matrix element where the dipoles have been subtracted.
The dotted line is the contribution from the finite parts of the integrated
dipoles. The solid red line is the sum of the two contributions. As this is by no means
a physical meaningful quantity we do not give actual numbers for the cross section. For the
sum, the error bars of the Monte Carlo integration are added.}
\end{figure}

Several checks have been performed to test the correctness of the different parts of the
implementation. A first check of the unintegrated dipoles is of course to check the limits, i.e.
the fact that the ratio of real emission matrix element over the sum of all subtraction terms,
\begin{equation}
 \frac{|{\cal M}_{real}|^2}{\sum {\cal D}} \to 1
\end{equation}
when one approaches a soft- and/or collinear limit. This could easily be done using the already
existing check program which is part of MadDipole (see \cite{Frederix:2008hu} for details).\\
This alone is not a sufficient test. Furthermore we compared results for single phase space points
against existing results for the processes $e^{+}e^{-}\to 4f$ with RacoonWW 
\cite{Denner:2002cg,Denner:1999gp,Denner:2000bj}, $e^{-}\gamma \to e^{-} \mu^{+} \mu^{-}$
\cite{Dittmaier:2008md} and $e^{+}e^{-}\to t \bar{t} H$ \cite{Denner:2003ri,Denner:2003zp}.
Thus all different splittings could be tested. This has been done for both
the unintegrated dipoles as well as for the integrated ones.\\ 
The aim of this package is to provide the necessary subtraction terms for single phase space
points without the phase space integration. In particular the user is not restricted to
MadEvent for the integration but can use MadDipole as a stand alone package and implement it
in a completely different framework. A pointwise agreement with existing results should therefore 
be considered sufficient to test the implementation. The calculation of complete cross sections 
(with virtual corrections from other packages) is beyond the scope of this work.\\ 
Furthermore we repeated the checks on the dependence of the cut parameter $\alpha$ as described
in \cite{Frederix:2010cj}. Although already being checked in the QCD version of MadDipole additional
checks were necessary because of two reasons. First of all the final-final dipole 
splitting function $f^{*}\to f\gamma$
is conventionally normalized differently  than in QCD, 
which made it necessary to calculate the correction terms (see appendix).
On the other hand having leptons in the initial state made it necessary to incorporate the concept
of calculating amplitudes at different points ($x=1$ and $x\ne1$) which is not needed for QCD processes.
Checking the independence of the total result on the 
phase space cut parameter provides a valuable test for
this part of the implementation.\\
In Fig.\ref{alphadep} we show the dependence of the $\alpha$-parameter for two processes,
$e^{+}e^{-}\to u \bar{u} \gamma$ and $e^{+}e^{-}\to t \bar{t} \gamma$. These two processes
contain the final-final splittings different from QCD and they also contain initial state radiation
off leptons. What the plots show is the cross section as a function of the cut parameter.
Actually, each splitting (final-final, final-initial, initial-final and initial-initial) has its
own $\alpha$-parameter which can be adjusted independently from each other. In this plot we change
them simultaneously so that their values are always equal.\\
The dashed line  shows the
cross section of the real emission matrix element where the dipoles have been subtracted so that
the difference is finite. The  dotted line shows the cross section of the finite parts 
of the integrated dipoles. Summing both contributions leads to the solid red line which shows
an independence on the cut parameter to a very good accuracy over several orders of magnitude.

\psfrag{alpha}{$\alpha$}
\psfrag{sigma}[][][1][-90]{$\sigma$}
\psfrag{TITLE}{\qquad \qquad
  \qquad $e^{+}\ e^{-} \rightarrow u\ \bar{u}\ \gamma$ }
\psfrag{TITLE1}{\qquad \qquad
  \qquad $e^{+}\ e^{-} \rightarrow t\ \bar{t}\ \gamma$ }

\section{Conclusions}
The MadDipole package automatically generates unintegrated and 
integrated subtraction terms required for NLO calculations. Previously, 
it provided all dipole subtraction terms for QCD 
radiation~\cite{Frederix:2008hu,Frederix:2010cj}. In this paper, we
extended MadDipole to include photonic radiation. The implementation is 
provided both in dimensional regularization and mass regularization. 
We include counterterms for the QED corrections to the parton distribution 
functions, and allow for subtraction terms for non-collinear-safe 
observables, i.e.\ with identified final state photons. 

With the photonic radiation now included in MadDipole, the MadGraph/MadEvent
framework can now be used for electroweak real radiation corrections
to collider observables. This development provides an important step 
towards the full automation of electroweak NLO corrections.  

\section*{Acknowledgments}
We are very grateful to Stefan Dittmaier and Marcus Weber for numerous valuable discussions,
support, providing results and performing comparisons. We also thank Rikkert Frederix and Tim Stelzer for their
help to make the code public.
This work was supported by the U.S.~Department of Energy under contract
No.~DE-FG02-91ER40677 and by the Swiss National Science
Foundation (SNF) under contract
200020-126691.

\begin{appendix}
 \section{Final-final $f\to f+\gamma$ dipole in dimensional regularization}
 The final-final dipole factor for the radiation of a photon off a fermion is 
 conventionally normalized differently from the corresponding QCD case.
 It is therefore documented in detail here. 
 For the case where a photon is radiated from a massless fermion
the auxiliary function is given by
(Eqn. (3.1) of  \cite{Dittmaier:1999mb})
\begin{eqnarray}
 \gsub_{ij,+}(p_i,p_j,k) &=&
\frac{1}{(p_i k)(1-y_{ij})} \biggl[
\frac{2}{1-z_{ij}(1-y_{ij})}-1-z_{ij} \biggr],
\nn\\[.5em]
\gsub_{ij,-}(p_i,p_j,k) &=& 0\;.
\end{eqnarray}
In this notation $k$ is the momentum of the photon, $p_i$ and $p_j$ are the momenta
of emitter and spectator. The variables $y_{ij}$ and $z_{ij}$ correspond to the
variables $y$ and $z$ of the QCD case. Note that there is an additional factor of
$\frac{1}{1-y_{ij}}$ present in this case.\\
The integrated term in dimensional regularization is obtained along the lines of the 
Catani-Seymour approach \cite{Catani:1996vz}.

We want to use dimensional regluarisation therefore we
are now following the Catani-Seymour approach. In that language the integrated
term is given by Eqns. (5.27,5.28) of \cite{Catani:1996vz}:
\begin{equation}
 {\cal V}_{ij,k} =
\int \left[dp_i({\widetilde p}_{ij},{\widetilde p}_k)\right]
\; \frac{1}{2 p_i \cdot p_j} \; <{\bom V}_{ij,k}>
\;\equiv \;\frac{\as}{2\pi}
\frac{1}{\Gamma(1-\ep)} \left(
\frac{4\pi \mu^2}{2 {\widetilde p}_{ij}{\widetilde p}_k} \right)^{\ep}
{\cal V}_{ij}(\ep) \;,
\end{equation}
and
\begin{equation}
{\cal V}_{ij}(\ep) =
\int_0^1 d{\tilde z}_i \,
\left({\tilde z}_i (1-{\tilde z}_i) \right)^{-\ep}
\int_0^1 \frac{dy}{y} \, \left( 1-y \right)^{1-2\ep} y^{-\ep} \;
\frac{<{\bom V}_{ij,k}({\tilde z}_i;y)>}{8 {\pi} \as
\mu^{2\ep}} \;.
\end{equation}
The transition can  be made
by replacing 
\begin{equation}
 \frac{<{\bom V}_{ij,k}({\tilde z}_i;y)>}{8 {\pi} \as
\mu^{2\ep}} \to 
\frac{1}{(1-y_{ij})} \biggl[
\frac{2}{1-z_{ij}(1-y_{ij})}-1-z_{ij}-\eps(1-z_{ij}) \biggr]\; .
\end{equation}
After integration and expanding in $\epsilon$ we find
\begin{equation}
 {\cal V}_{\gamma f}(\eps)=\frac{1}{\epsilon ^2}+\frac{3}{2 \epsilon }+\left(\frac{7}{2}-\frac{\pi
   ^2}{6}\right)+O\left(\epsilon ^1\right)\;.
\end{equation}
The massive case is more involved. We decompose the integral into an eikonal and
a collinear part,
\begin{equation}
I_{\gamma f,k}(\mu_j,\mu_k;\eps)  =
\left[ 2I^{\eik}(\mu_j,\mu_k;\eps)
        + I^{\coll}_{\gamma f,k}(\mu_j,\mu_k;\eps) \right]\:,
\end{equation}
where we define the eikonal integral in analogy to Eq.\ (5.26) of \cite{Catani:2002hc}
\begin{equation}
 \frac{\alps}{2\pi}\frac{1}{\Gamma(1-\eps)}
\biggl(\frac{4\pi\mu^2}{Q^2}\biggr)^\eps I^{\eik}(\mu_j,\mu_k;\eps) =
\int [\rd p_i(\tpij,\tpk)] \, \frac{1}{2p_i p_j} \,\frac{1}{R_{ij}(y)}
\frac{8\pi\mu^{2\eps}\alps}{1-\zj(1-\yijk)}\;.
\end{equation}
The difference to the CS dipoles is given by the additional factor $R_{ij}(y)$,
which is defined in Eq.~(4.3) of  \cite{Dittmaier:1999mb}
\begin{equation}
 R_{ij}(y) = 
\frac{\sqrt{(2m_j^2+\bar P_{ij}^2-\bar P_{ij}^2 y)^2-4P_{ij}^2 m_j^2}}
{\sqrt{\lambda_{ij}}}, 
\end{equation}
with
\begin{equation}
 P_{ij} = p_i+p_j+k, \qquad \bar P_{ij}^2 = P_{ij}^2-m_i^2-m_j^2-m_\gamma^2,
\qquad \lambda_{ij}=\lambda(P_{ij}^2,m_i^2,m_j^2)\;.
\end{equation}
Performing the integration we find for the eikonal integral
\begin{eqnarray}
 I^{\eik}&(\mu_j,\mu_k;\eps)=\frac{1}{\epsilon}\frac{\ln \rho_j}{\tilde{v}}+
 \frac{1}{\tilde{v}}\left\{ \text{Li}_2\left(\frac{(c_j-\tilde{v}+1) (c_k
 -\tilde{v}+1)}{(c_j+c_k+2) (c_k-y_{+}+1)-c_2}\right) +
   \text{Li}_2\left(\frac{(c_j+c_k+2) (c_k-y_{+}+1)-c_2}{(c+\tilde{v}+1)(c_k+\tilde{v}+1)}
  \right) \right.\nn \\&\left.
 + 2\ln (\rho)\ln ((c_j+c_k+2) (c_k-y_{+}+1)-c_2)+\frac{1}{2}
  \ln ^2\left(\frac{(c_j+c_k+2) (-c_k+y_{+}-1)+c_2}{(c_j-\tilde{v}+1)(-c_k+\tilde{v}-1)}\right)\right.\nn\\&\left.
 + \ln (2\tilde{v}) \ln \left(\frac{(c_j+c_k+2) (c_k-\tilde{v}+1)-c_2}{c_j-\tilde{v}+1}\right)+
   \left(\ln (c_j-\tilde{v}+1) \ln
   \left(\frac{c_k+\tilde{v}+1}{\rho_j^2}\right)\right.\right.\nn\\& \left.\left.
 -\ln \left(2 \tilde{v} (c_j-\tilde{v}+1)\right)\ln (c_k-\tilde{v}+1)-\ln \left(2\tilde{v}\right)
 \ln (c_j-\tilde{v}+1)\right)+
 \text{Li}_2\left(\frac{c_j-\tilde{v}+1}{c_j+y_{+}+1}\right)\right.\nn\\&\left.
 +\text{Li}_2\left(\frac{c_j+y_{+}+1}{c_j+\tilde{v}+1}\right)+\ln (c_j+\tilde{v}+1)
 \ln \left(\frac{2 \tilde{v}}{\rho_j^2}\right)+
 2\ln (\rho_j) \ln(c_j+y_{+}+1)\right.\nn\\&\left.
+\frac{1}{2} \ln ^2\left(\frac{c_j+y_{+}+1}{c_j-\tilde{v}+1}\right)- \text{Li}_2\left(\frac{c_k-\tilde{v}+1}{c_k-y_{+}+1}\right)
- \text{Li}_2\left(\frac{c_k-y_{+}+1}{c_k+\tilde{v}+1}\right)- 2\ln \left(\rho_k\right) \ln
   (c_k-y_{+}+1)\right.\nn\\&\left.
+ \ln \left(\rho_j\right) \ln \left(\frac{(\tilde{v}-y_{+})^2 
 (\tilde{v}+y_{+})^2}{4 \rho_j^2 \tilde{v}^2
   (c_k-y_{+}+1)^2}\right)-\frac{1}{2} \ln ^2\left(\frac{c_k-y_{+}+1}{c_k-\tilde{v}+1}\right)
   - \text{Li}_2(\rho^2)+ \text{Li}_2(\rho_k^2)\right.\nn\\&\left.
- \text{Li}_2(\rho_j^2)-\frac{\pi ^2}{6} +\text{Li}_2\left(-\frac{2 \tilde{v}}{\rho_j^2
   (c_j+\tilde{v}+1)}\right)-\ln (\rho_j) \ln \left(\frac{2 y_{+}^2 \sqrt{\lambda }}
 {c \rho_j \tilde{v}}\right) \right\}+{\cal O}(\epsilon)\;,
\end{eqnarray}
where we used the following abbreviations:
\begin{eqnarray}
 \sqrt{\lambda}&=\sqrt{\lambda(1,\mu_j^2,\mu_k^2)}\quad, 
\rho_n(\mu_j,\mu_k) = \sqrt{
\frac{1-\tvijk+2\mu_n^2/(1-\mu_j^2-\mu_k^2)}
     {1+\tvijk+2\mu_n^2/(1-\mu_j^2-\mu_k^2)}} \quad (n=j,k) \;,\nn\\
\tilde{v}&=\tilde{v}_{ij,k}, \hspace{3cm} \rho = \sqrt{\frac{1-\tvijk}{1+\tvijk}},\hspace{5cm}
\end{eqnarray}
and
\begin{equation}
 y_+ = 1-\frac{2\mu_k(1-\mu_k)}{1-\mu_i^2-\mu_j^2-\mu_k^2}
\end{equation}
is the upper limit of the $y$ integration. Furthermore we define for brevity
\begin{equation}
 c_j=\frac{2\mu_j^2}{1-\mu_j^2-\mu_k^2},\quad c_k=\frac{2\mu_k^2}{1-\mu_j^2-\mu_k^2},
\quad c_2=\frac{4\mu_k^2}{(1-\mu_j^2-\mu_k^2)^2}\;.
\end{equation}
The contribution from the collinear integral is simpler and we find
\begin{eqnarray}
  I^{\coll}_{\gamma f,k}&(\mu_f,\mu_k;\eps)=\frac{1}{\epsilon}+\frac{1}{2} \left(-4 \log \left(-\mu_j^2+\mu_k^2-2 \mu_k+1\right) \right.\nn \\& \left.
+\frac{\mu_j^2}{(\mu_k-1)^2}+3\right)+\log (\mu_j(1 - \mu_k))+{\cal O}(\epsilon)\;.
\end{eqnarray}
The results in dimensional regularization can be compared with the results in mass regularization given in Eq. (4.10) of 
\cite{Dittmaier:1999mb}. Comparing the two results numerically we find full agreement.
\\ \\
As the limit $\epsilon \to 0$ does not commute with the limit of vanishing masses
we also give the explicit result for the case when one mass is set to zero. In the case of
a vanishing emitter mass we find
\begin{eqnarray}
 I^{\eik}&(0,\mu_k;\eps)=\frac{1}{2\epsilon^2}-\frac{\ln(1-\mu_k^2)}{\epsilon}
+2\text{Li}_2\left(\frac{1}{\mu_k+1}\right)+\ln \left(\frac{1}{\mu_k}-1\right) \ln (1-\mu_k)\nn\\&+\ln
   (\mu_k+1) (4 \ln (1-\mu_k)-3 \ln (\mu_k)+2 \ln (\mu_k+1))-\frac{7 \pi ^2}{12}+{\cal O}(\epsilon)\;,
\end{eqnarray}
whereas for vanishing spectator mass we find
\begin{flalign}
 I^{\eik}(\mu_j,0;\eps)&=\frac{\ln(\mu_j)}{\epsilon}+
2 \text{Li}_2\left(1-\frac{1}{\mu_j^2}\right)-\text{Li}_2\left(\mu_j^2\right)-4 \log
   \left(1-\mu_j^2\right) \log (\mu_j)\nn\\ &+3 \log ^2(\mu_j)+\frac{\pi ^2}{6}+{\cal O}(\epsilon)\;.
\end{flalign}
The corresponding terms from the collinear integral read as
\begin{equation}
I^{\coll}_{\gamma f,k}(0,\mu_k;\eps)=\frac{3}{2\epsilon}+\frac{7}{2}-3 \log (1-\mu_k)+{\cal O}(\epsilon)\;,
\end{equation}
and
\begin{equation}
I^{\coll}_{\gamma f,k}(\mu_j,0;\eps)=\frac{1}{\epsilon}+\frac{1}{2} \left\{\mu_j^2
-4 \log \left(1-\mu_j^2\right)+2 \log (\mu_j)+3\right\}+{\cal O}(\epsilon)\;. 
\end{equation}

\subsection{$\alpha$-cut parameter}
The idea and the use of the $\alpha$-parameter is described in detail in \cite{Nagy:1998bb,Nagy:2003tz}.
As the final-final splitting function for the splitting $f^*\to f \gamma$ is different from 
the one in QCD the results were not present in MadDipole so far.\\
These correction terms only affect finite terms and have no influence on singular terms which means
they can be calculated either in the mass regularization approach or in the dimensional regularization
approach. The correction terms are the same so one can use the method that appears to be more suitable.\\
The general form of the correction terms in the language of dimensional regularization is given by
\begin{equation}
 I_{\gamma f,k}(\mu_j,\mu_k; \epsilon, \alpha)=I_{\gamma f,k}(\mu_j,\mu_k; \epsilon) 
+ \Delta I_{\gamma f,k}(\mu_j,\mu_k; \alpha)\;.
\end{equation}
The massless case is straightforward and we find
\begin{equation}
 \Delta I_{\gamma f,k}(0,0; \alpha) =-2 \text{Li}_2(1-\alpha )-\log ^2(\alpha )-\frac{3 \log (\alpha )}{2}\; .
\end{equation}
In the massive case we calculate eikonal integral and collinear integral seperately.
For the case where both emitter and spectator are massive we find
\begin{flalign}
 \Delta I&^{\eik}(\mu_j,  \mu_k,\alpha)=\frac{1}{2v}\left \{2 \text{Li}_2
  \left(\frac{v^2+c+c c_k+c_k-(c+c_k+2) y_{\alpha}+1}{(c+v+1) (c_k+v+1)}\right)
\right. \nn \\& \left.
-2\text{Li}_2\left(\frac{v^2+c+c c_k+c_k-(c+c_k+2) y_{+}+1}{(c+v+1) (c_k+v+1)}\right)+2
   \text{Li}_2\left(\frac{(c-v+1) (-c_k+v-1)}{(c+c_k+2) (y_{\alpha}-1)}\right)
\right. \nn \\& \left.
-2 \text{Li}_2\left(\frac{(c-v+1)
   (-c_k+v-1)}{(c+c_k+2) (y_{+}-1)}\right)
-2\text{Li}_2\left(\frac{c-v+1}{c+y_{+}+1}\right)-2 \text{Li}_2\left(\frac{c+y_{+}+1}{c+v+1}\right)
\right. \nn \\& \left.
-2 \log \left(\frac{v+y_{\alpha}}{v-y_{\alpha}}\right) \log
   \left(\frac{(c+y_{\alpha}+1) (c_k-y_{\alpha}+1)}{-y_{\alpha} (c+c_k+2)+c c_k+c+c_k+v^2+1}\right)
\right. \nn \\& \left.
+2 \log
   \left(\frac{v+y_{+}}{v-y_{+}}\right) \log \left(\frac{(c+y_{+}+1) (c_k-y_{+}+1)}{-y_{+}
   (c+c_k+2)+c c_k+c+c_k+v^2+1}\right)
\right. \nn \\& \left.
+(\log (1-y_{\alpha})-\log (1-y_{+})) (-2 \log ((c+v+1)
   (c_k+v+1))+2 \log (c+c_k+2)
\right. \nn \\& \left.
+\log (1-y_{\alpha})+\log (1-y_{+}))+2
   \text{Li}_2\left(\frac{c-v+1}{c+y_{\alpha}+1}\right)+2 \text{Li}_2\left(\frac{c+y_{\alpha}+1}{c+v+1}\right)
\right. \nn \\& \left.
-(\log
   (c+y_{\alpha}+1)-\log (c+y_{+}+1)) (2 \log (c+v+1)-\log (c+y_{\alpha}+1)-\log (c+y_{+}+1))
\right. \nn \\& \left.
+(\log (c_k-y_{\alpha}+1)-\log
   (c_k-y_{+}+1)) (2 \log (c_k+v+1)
-\log (c_k-y_{\alpha}+1)-\log (c_k-y_{+}+1))
\right. \nn \\& \left.
-2
   \text{Li}_2\left(\frac{c_k-v+1}{c_k-y_{\alpha}+1}\right)-2
   \text{Li}_2\left(\frac{c_k-y_{\alpha}+1}{c_k+v+1}\right)+2
   \text{Li}_2\left(\frac{c_k-v+1}{c_k-y_{+}+1}\right)+2
   \text{Li}_2\left(\frac{c_k-y_{+}+1}{c_k+v+1}\right)
\right\},
\end{flalign}
with
\begin{equation}
 y_{\alpha}=\sqrt{\left(c_k (1-\alpha )+\left(\sqrt{c_2}-1\right) \alpha 
+1\right)^2-c_2}+ \alpha(c_k-\sqrt{c_2} +1)\;.
\end{equation}
\\
For a massless emitter and a massive spectator the correction term is given by
\begin{flalign}
 \Delta I^{\eik}(0,& \mu_k,\alpha)=
\log (1-y_{\alpha}) \log \left(-\frac{4}{\left(\mu_k^2-1\right) (y_{\alpha}+1)^2}\right)\log (y_{\alpha}+1)
   \log \left(-\left(\mu_k^2-1\right) (y_{\alpha}+1)\right)\nn\\&+2
   \text{Li}_2\left(\frac{1}{\mu_k+1}\right)-\log (2-2 \mu_k) \log
   \left(\frac{2}{\mu_k+1}\right)+\log \left(\frac{2 \mu_k}{\mu_k+1}\right) \log
   \left(\frac{2}{\mu_k+1}-1\right)\nn\\&-2 \text{Li}_2\left(\frac{y_{\alpha}+1}{2}\right),
\end{flalign}
with
\begin{equation}
 y_{\alpha}=\frac{ \alpha(1-\mu_k) +\sqrt{(\alpha -1) \left((\mu_k-1)^2 \alpha -(\mu_k+1)^2\right)}
   }{\mu_k+1}\;.
\end{equation}
\\
Finally for the case of a massive emitter and a massless spectator we find
\begin{flalign}
 \Delta I^{\eik}(\mu_j,& 0,\alpha)=
-\text{Li}_2\left(\alpha -\frac{\alpha }{\mu_j^2}\right)-\text{Li}_2\left(-\alpha  \mu_j^2+\mu_j^2+\alpha
   \right)+\text{Li}_2\left(1-\frac{1}{\mu_j^2}\right)\nn\\&-\log \left(\left(\mu_j^2-1\right) (\alpha -1)\right)
   \log \left(\mu_j^2 (-\alpha )+\mu_j^2+\alpha \right)+2 \log (\mu_j) \log (\alpha )+\frac{\pi ^2}{6}.
\end{flalign}
In the same way as for the eikonal integral we take the $\alpha$-dependence of the collinear integral into account by
writing
\begin{equation}
 I_{\gamma f,k}^{coll}(\mu_j,\mu_k;\epsilon,\alpha)=I_{\gamma f,k}^{coll}(\mu_j,\mu_k;\epsilon)+\Delta I_{\gamma f,k}^{coll}(\mu_j,\mu_k;\alpha).
\end{equation}
In the general case where both emitter and spectator are massive the correction term reads as
\begin{flalign}
 \Delta I_{\gamma f,k}^{coll}(\mu_j,\mu_k;\alpha)&=-\frac{2 (\alpha -1) c_j y_{+}+(c_j+2 y_{+}) (2 \alpha 
 y_{+}+c_j) \left(\log \left(\frac{c_j+2 y_{+}}{2 \alpha  y_{+}+c_j}\right)+4 \log
   (\alpha )\right)}{2 (c_j+2 y_{+}) (2 \alpha  y_{+}+c_j)}\;.
\end{flalign}
The limits $\mu_j \to 0$ and $\mu_k \to 0$ can  be performed easily and one finds
\begin{flalign}
 \Delta I_{\gamma f,k}^{coll}(\mu_j,0;\alpha)&=-\frac{1}{2} \left\{\frac{(\alpha -1) \left(\mu_j^2-1\right) 
\mu_j^2}{(\alpha -1) \mu_j^2-\alpha }-\log \left(-\alpha 
   \mu_j^2+\alpha +\mu_j^2\right)+4 \log (\alpha )\right\}\;,
\end{flalign}
and
\begin{flalign}
 \Delta I_{\gamma f,k}^{coll}(0,\mu_k;\alpha)=-\frac{3}{2}\log(\alpha)
\end{flalign}
respectively.\\
For these calculations we have used the \texttt{HypExp} package for Mathematica \cite{Huber:2007dx,Huber:2005yg}.\\
As mentioned above the $\alpha$-terms can as well be calculated using the formalism of mass regularization.
To check our result we performed a second calculation using mass regularization and making use of the formulae
giving in the appendix of \cite{Dittmaier:1999mb} we found agreement between the two approaches.  \\

\end{appendix}

\bibliographystyle{utphys}
\bibliography{references}
\end{document}